\PassOptionsToPackage{colorlinks, linkcolor=blue, citecolor=blue, urlcolor=blue}{hyperref}
\documentclass{article}

\usepackage{jheppub}

\usepackage{amsmath}
\usepackage{amssymb}
\usepackage[capitalize,nameinlink]{cleveref}
\usepackage{graphicx}
\usepackage{hyperref}
\usepackage[dvipsnames]{xcolor}
\usepackage{xspace}
\usepackage{tikz}

\usetikzlibrary{decorations.pathmorphing}

\newcommand{\EOS}{\texttt{EOS}\xspace}
\newcommand{\dynesty}{\texttt{dynesty}\xspace}
\newcommand{\kperp}{\ensuremath{k_\perp}\xspace}
\newcommand{\kperpth}{\ensuremath{k_\perp^\text{th}}\xspace}
\newcommand{\kperpdet}{\ensuremath{k_\perp^\text{det}}\xspace}
\newcommand{\GeV}{\text{GeV}}
\newcommand{\ie}{\textit{i.e.}\xspace}
\newcommand{\eg}{\textit{e.g.}\xspace}

\title{Exploiting Perpendicular Momentum Distributions of Semileptonic Decays: $\bar{B}_s^0\to D_s^+\mu^-\bar\nu$ as a Case Study}
\author[a]{Charles Earnshaw,}
\emailAdd{charlie.earnshaw@gmail.com}
\author[b]{Biljana Mitreska,}
\emailAdd{biljana.mitreska@cern.ch} 
\author[a]{Danny van Dyk}
\emailAdd{danny.van.dyk@gmail.com}
\affiliation[a]{%
    Institute for Particle Physics Phenomenology and Department of Physics, Durham University,\\
    Durham DH1~3LE, UK
}
\affiliation[b]{%
    Department of Physics and Astronomy, University of Manchester,\\
    Manchester M13 9PL, UK
}
\date{October 2025}

\begin{document}

\abstract{%
    We derive the differential distribution of semileptonic decays
    with respect to the perpendicular momentum component of the
    final state hadron.
    The benefits and shortfalls arising from measurements
    of these distributions are discussed.
    Our approach is illustrated on the LHCb measurement of the $\bar{B}_s^0\to D_s^+\mu^-\bar\nu$
    decay distribution where the publicly available data by the LHCb experiment is used in an independent phenomenological
    analysis for the first time.
    We extract the CKM element $|V_{cb}|$ and information on the shape of the
    relevant hadronic form factors from the measurement of the binned rate in the perpendicular momentum component of the hadron.
}

\begin{flushright}
    EOS-2025-06\\
    IPPP/25/83
\end{flushright}
\vspace*{-3\baselineskip}

\maketitle

\section{Introduction}
\label{sec:intro}

The Cabibbo-Kobayashi-Maskawa quark mixing matrix~\cite{Cabibbo:1963yz,Kobayashi:1973fv} is at the core of the
Standard Model (SM) of particle physics. Within the SM paradigm, it
describes the misalignment between flavour eigenstates and mass eigenstates
as a unitary $3\times 3$ matrix.
Its matrix elements can be expressed in terms of three
real-valued parameters and one complex phase.
The SM does not predict these parameters, and their determination from global fits is
key to understanding the flavour structure of the SM~\cite{Charles:2004jd,UTfit:2006vpt}.
Central to determining the CKM matrix elements from data are semileptonic decays,
both in terms of measurements of the branching ratio as well as in terms
of theoretical predictions.\\

Beyond the SM, new interactions and new matter particles are theorised to
exist. Their effects can modify the short-distance coupling strength
that is determined in CKM analysis of semileptonic decays.
These modifications have the potential to affect the overall normalisation
of the decay rate and the kinematical distribution of the decay products.
However, both types of effects can also be caused by the mismodelling of the
SM decay rate, specifically, the mismodelling of the relevant hadronic matrix elements.
We can differentiate these effects within global analyses of semileptonic decays,
\ie, statistical analyses that overconstrain the short-distance parameters by
using data stemming from multiple different decay processes into account.
The purpose of this article is to demonstrate this and to explain how a new type of measurement can be used in such global analyses.\\

Performing experimental measurements with $b \to c \ell \nu$ transitions is challenging due to the presence of the neutrino in the final state~\cite{Gambino:2020jvv}.
The B factory experiments BaBar, Belle, and Belle II, together with the BEPC experiments BES-I to BES-III,
the CLEO experiment, and the KLOE experiment, have an excellent track record
in measuring absolute branching fractions of semileptonic decays; see~\cite{%
BaBar:2014omp,%
Belle-II:2018jsg,%
BESIII:2024slx,%
CLEO:2010enr,%
KLOE:2007jte%
} for representative reviews or examples covering each of these experiments.
This is due to the fact that all of these experiments benefit from
known initial state kinematics, which allows for reconstructing or constraining
the neutrino momentum.
In contrast, being run at a $pp$ machine, the LHCb experiment does not incur the same benefit
and, moreover, requires branching ratio measurements to be performed relative to external measurements of the branching ratio for a reference decay channel.
Nevertheless, LHCb has made important measurements of semileptonic
decays using non-trivial methods that account for the neutrino reconstruction~\cite{LHCb:2023zxo,LHCb:2023uiv,LHCb:2024jll}. These measurements involve
solving a nonlinear equation for $q^2$, the mass squared of the lepton-neutrino system, using the kinematic variables of the detectable particles as inputs.
This method yields two solutions that cannot be distinguished on an event-by-event basis~\cite{Dambach:2006ha}.
One may select between them using different strategies; these include, for example, choosing randomly,
or using multivariate regression algorithms to inform the choice~\cite{Ciezarek:2016lqu} to ensure that \emph{on average} the correct solution is chosen more frequently than not.
In the case of $\bar{B}_s^0\to K^+\mu^-\bar\nu$ the method proposed in Ref.~\cite{Ciezarek:2016lqu} selects the correct solution at a rate of about $70\%$.\\

\begin{figure}[t]
    \centering
    \begin{tikzpicture}[scale=1.25]
        \fill (0,0) circle (0.7mm); 

        \draw[dashed] (-4.5,0) --  (4.5,0) node[below right] {\(\hat{z}_B\)};

        \draw[thick] (0,0) --  (4, 2) node[near end, below] {\(D_s^+\)};

        \draw[-latex] (0.5,0.5) --  (2.5,1.5) node[midway, above] {$k$};

        \draw[dotted] (4,2) -- (4,0);

        \draw[-latex] (4.25,0.25) -- (4.25,1.75) node[midway, right] {\(k_\perp\)};

        \draw[decorate, decoration={snake, amplitude=1mm, segment length=2.5mm}] (0,0) -- (-4,-2) node[near end,above, xshift = 3pt,yshift = 3pt] {\(W^*\)};

        \draw[-latex] (-0.5,-0.6) --  (-2.5,-1.6) node[midway, below] {\(q\)};

        \draw[dotted] (-4,-2) -- (-4,0);

        \draw[-latex] (-4.25,-0.25) -- (-4.25,-1.75) node[midway, left] {\(k_\perp\)};

        \draw[thick] (0,0) ++ (0:1) arc (0:27:1) node[midway, right] {\(\theta_B\)};
        \draw[thick] (0,0) ++ (180:1) arc (180:205:1) node[midway, left] {\(\theta_B\)};

        \draw[-latex] (4.5,0) ++ (270:0.75) arc (270:360:0.75) node[midway, right] {\(\phi_B\)};
    \end{tikzpicture}
    \caption{%
        Sketch of the decay kinematics and illustration of the
        $k_\perp$ variable in the rest frame of the $\bar{B}_{s}^0$ meson.
        The momentum of the $D_{s}^+$ meson is labelled $k$,
        and the momentum transferred to the lepton-neutrino system
        by means of a virtual $W^*$ boson is labelled $q$.
        The $\hat{z}_B$ direction corresponds to the direction
        of flight of the $\bar{B}_{s}^0$ meson in the laboratory frame.
        The solid angle between the axis $D_{s}^+$--$W^*$ and
        the $\hat{z}_B$ axis is described by the azimuthal angle
        $\theta_B$ and the polar angle $\phi_B$.
    }
    \label{fig:intro:sketch}
\end{figure}
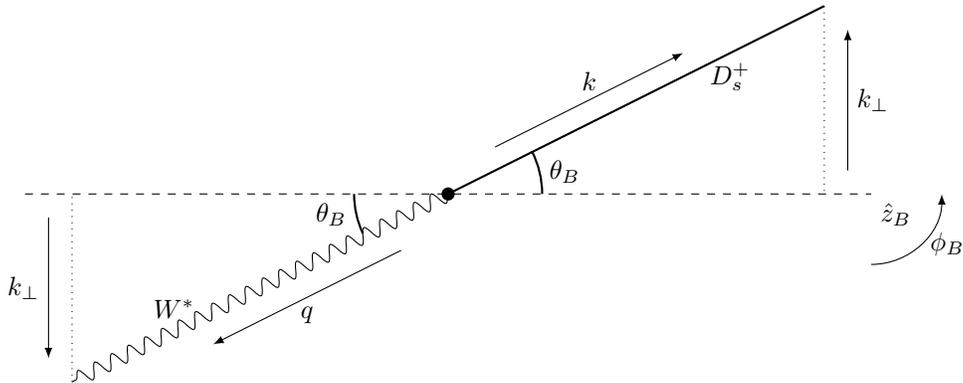

As an alternative to the above approach, the LHCb collaboration has piloted a measurement
of the decay $\bar{B}_s^0(p)\to D_s^{+}(k) \mu^-(q_1)\bar\nu(q_2)$ (see \cref{fig:intro:sketch} for parts of the decay topology)
in which a different approach to handling the missing neutrino is pursued~\cite{LHCb:2020cyw}.
In that measurement, the distribution of the decay rate with respect to a different variable $k_\perp$, rather than $q^2$, is determined.
In this work,\footnote{%
    In Ref.~\cite{LHCb:2020cyw}, the $D^+_s$ meson momentum is labelled $p$ and
    the variable in question is hence labelled $p_\perp$.
} $k_\perp$ is the projection component of the spatial momentum $\vec{k}$ onto the direction perpendicular to the $\Bar{B}^0_s$-meson
direction of flight. These definitions are illustrated in \cref{fig:intro:sketch}.
This variable can be determined unambiguously from the visible
kinematics. However, it does not unambiguously relate to $q^2$, as we will show later.
The purpose of this article is to document how
to use forward modelling to translate the $q^2$ into the $\kperp$ rate to exploit measurements of the latter
in phenomenological analyses.\\

The structure of this article is as follows.
We begin in \cref{sec:theoretical-predictions} with deriving theory predictions for
the decay distribution in terms of $k_\perp$ by using predictions
for the decay distribution in terms of $q^2$.
In \cref{sec:detector-effects}, we continue by discussing detector
effects affecting the measurement of the $k_\perp$ distribution
published in Ref.~\cite{LHCb:2020cyw}.
In \cref{sec:pheno}, we use the results of the two preceding sections
to perform the first phenomenological analysis of the measurements published in Ref.~\cite{LHCb:2020cyw}
outside of the LHCb experiment.
We summarise our findings in \cref{sec:summary}.
In \cref{app:other-variables}, we further illustrate that the decay distribution with
respect to the perpendicular component of any other of the
visible final-state momenta cannot be determined unambiguously from
the kinematic variable available to the theoretical predictions.

\section{Theoretical predictions}
\label{sec:theoretical-predictions}

In this section, we derive the kinematic decay distribution of a semileptonic decay as a function
of the kinematic variable \kperp. Although our results can be used in the context of any semileptonic decay of
a pseudoscalar meson without change, we notate them for the decay $\bar{B}_s^0\to D_s^+\mu^-\bar\nu$,
in anticipation of applications later on.\\

For predictions of semileptonic decays such as
$\bar{B}_s^0(p) \to D_s^+(k) \mu^-(q_1)\bar\nu(q_2)$,
it is common to discuss the theoretical distribution in terms of either
$q^2 \equiv (q_1 + q_2)^2$ --- the mass squared of the lepton-neutrino system --- 
or $w \equiv (M_{B_{s}}^2 + M_{D_s}^2 - q^2) / (2 M_{B_s} M_{D_s})$
 --- the recoil of the child meson in the rest frame of the parent meson.
Throughout this work, we use the description in terms of $q^2$.\\

In the SM and for massless leptons, the differential branching ratio
for the decay at hand reads
\begin{equation}
    \label{eq:theoretical-predictions:differential-br}
    \frac{d\mathcal{B}(\bar{B}_s^0\to D_s^+\mu^-\bar\nu)}{dq^2}
        = \frac{G_\text{F}^2 M_{D_s}^2 \lambda(M_{B_s}^2, M_{D_s}^2, q^2)}{48 \pi^3 M_{B_s}} \, |V_{cb}|^2 \, |f_+(q^2)|^2\,.
\end{equation}
Here $G_\text{F}$ is the Fermi constant, $\lambda(a, b, c) = a^2 + b^2 + c^2 - 2ab - 2ac - 2bc$
denotes the K\"all\'en function, and $f_+(q^2)$ denotes
the vector form factor for $\bar{B}_s^0\to D_s^+$ transitions.
The above does not include contributions due to non-zero lepton mass and beyond-the-SM (BSM) effects;
see Ref.~\cite{Duraisamy:2014sna} for the expression
assuming general BSM contributions in the weak effective theory.\\

For this phenomenological analysis, it is sufficient to have access to the
differential branching fraction as a function of $q^2$ to derive the
differential branching fraction as a function of the perpendicular
momentum component \kperp.
For our derivation, we use the following conventions as sketched in \cref{fig:intro:sketch}.
We work in the rest frame of the $\bar{B}_s^0$ meson, as indicated by a $B$
subscript to the relevant quantities.
The $z$ axis of our coordinate system corresponds to the direction
of flight of the initial $\bar{B}_s^0$ meson in the laboratory frame, which we denote as $\hat{z}_B$.
The direction of flight of the $D_s^+$ meson is angled upward within
the $x$---$z$ plane by the azimuthal angle $\theta_B$. It is further
angled out of the $x$---$z$ plane into $y$ direction by the
polar angle $\phi_B$. The momentum \kperp describes the
projection onto a direction perpendicular to the $z$ axis.
In this frame, we obtain for the square of the perpendicular momentum
component $\kperp$
\begin{equation}
    \label{eq:theoretical-predictions:k_perp2}
    \kperp^2\big|_B \equiv \frac{\lambda(M_{B_s}^2, M_{D_s}^2, q^2)}{4 M_{B_s}^2} \sin^2 \theta_B\,.
\end{equation}
Note that there is no dependence on the polar angle $\phi_B$.\\

In the reconstruction of a single decay event in
the laboratory frame, the perpendicular
momentum is fully determined from the kinematic variables attributed
to ``visible'' states, \ie, the $\bar{B}^0_s$ flight direction ($\hat{z}_B$)
and the $D_s^+$ momentum ($k$). For any given signal event, it is not possible to unambiguously
relate the momentum from the laboratory frame to the $\bar{B}_s^0$ rest frame,
as discussed in the introduction and detailed in Ref.~\cite{Ciezarek:2016lqu}.
However, the laboratory frame and our choice of the $\bar{B}_s^0$ rest frame
are connected by a Lorentz boost along the $\hat{z}_B$ direction.
Such boosts leave the perpendicular components of all four vectors
invariant and hence ${\kperp}\big|_B = \kperp\big|_\text{lab}$.
It is therefore sufficient to calculate the differential branching ratio
$d\mathcal{B} / d\kperp$ in the $\bar{B}_s^0$ rest frame.\\

To do so, we start with the definition of the integrated branching ratio,
which is a Lorentz invariant,
\begin{equation}
    \mathcal{B}
        = \iint dq^2 d\!\cos\theta_B      \frac{d^2\mathcal{B}}{dq^2\, d\!\cos\theta_B}\\
        = \iint d\kperp^2 d\!\cos\theta_B \frac{d^2\mathcal{B}}{d\kperp^2\, d\!\cos\theta_B}\,.
\end{equation}
Next, we use that the initial $\bar{B}^0_s$ meson is a (pseudo)scalar state and therefore that the
solid angle $(\cos\theta_B, \phi_B)$ describing the direction of flight of the $D_s^+$ meson is
distributed isotropically.
As a consequence, in its rest frame, we obtain
\begin{equation}
    \label{eq:theoretical-predictions:isotropic}
    \frac{d^2\mathcal{B}}{dq^2\, d\!\cos\theta_B}
        = \mathcal{U}_{-1,+1}(\cos\theta_B) \frac{d\mathcal{B}}{dq^2} 
        = \frac{1}{2}\, \frac{d\mathcal{B}}{dq^2}\,,
\end{equation}
where $\mathcal{U}_{-1,+1}(\cos\theta_B) = 1/2$ denotes the uniform PDF of the azimuthal angle in its domain.
The generalisation to initial states with total angular momentum $J=1/2$ or $J=1$ requires knowledge of the
polarisation of the state to access the decay distribution, \eg using a density-matrix formalism.\\

From the above, we read off
\begin{equation}
    \label{eq:theoretical-predictions:differential-br-kperp2}
    \begin{aligned}
        \frac{d\mathcal{B}}{d\kperp^2}
            & = \int_{-1}^{+1} d\!\cos\theta_B \frac{d^2\mathcal{B}}{d\kperp^2\, d\!\cos\theta_B}\\
            & = \int_{-1}^{+1} d\!\cos\theta_B \left|\frac{\partial(q^2, \cos\theta_B)}{\partial(\kperp^2, \cos\theta_B)}\right|
                               \frac{d^2\mathcal{B}}{q^2\, d\!\cos\theta_B} \, \Theta(q^2 - m_\ell^2)\,
                               \bigg|_{q^2 = q^2(\kperp^2, \cos\theta_B)}\\
            & = \frac{1}{2} \int_{-1}^{+1} d\!\cos\theta_B \left|\frac{\partial(q^2, \cos\theta_B)}{\partial(\kperp^2, \cos\theta_B)}\right|
                               \frac{d\mathcal{B}}{q^2} \, \Theta(q^2 - m_\ell^2)\,
                               \bigg|_{q^2 = q^2(\kperp^2, \cos\theta_B)}
                               \,,
    \end{aligned}
\end{equation}
where $\Theta$ denotes the Heaviside function and we introduce the Jacobian
\begin{equation}
    \left|\frac{\partial(q^2, \cos\theta_B)}{\partial(\kperp^2, \cos\theta_B)}\right|
         = \left|\frac{M_{B_s}}{\left(1 - \cos^2{\theta_B}\right)\sqrt{M^2_{D_s}+\frac{\kperp^2}{1-\cos^2\theta_B}}}\right|\,.
\end{equation}
In \cref{eq:theoretical-predictions:differential-br-kperp2} we rewrite $q^2$ in terms of $\kperp^2$ and $\cos\theta_B$ by means of
\begin{equation}
    \label{eq:q2fromkperpandtheta}
    q^2(\kperp^2, \cos\theta_B) = M_{B_s}^2 + M_{D_s}^2
        - 2 M_{B_s} \sqrt{%
            \frac{\kperp^2}{1 - \cos^2\theta_B} + M_{D_s}^2
        }\,,
\end{equation}
which corresponds to the physical branch of solving \cref{eq:theoretical-predictions:k_perp2} for $q^2$.
For the comparison with the experimentally determined rate in the variable $\kperp$, one further
Jacobian $\partial \kperp^2 / \partial \kperp = 2\kperp$ is needed.\\

\begin{figure}
    \centering
    \includegraphics[width=0.6\linewidth]{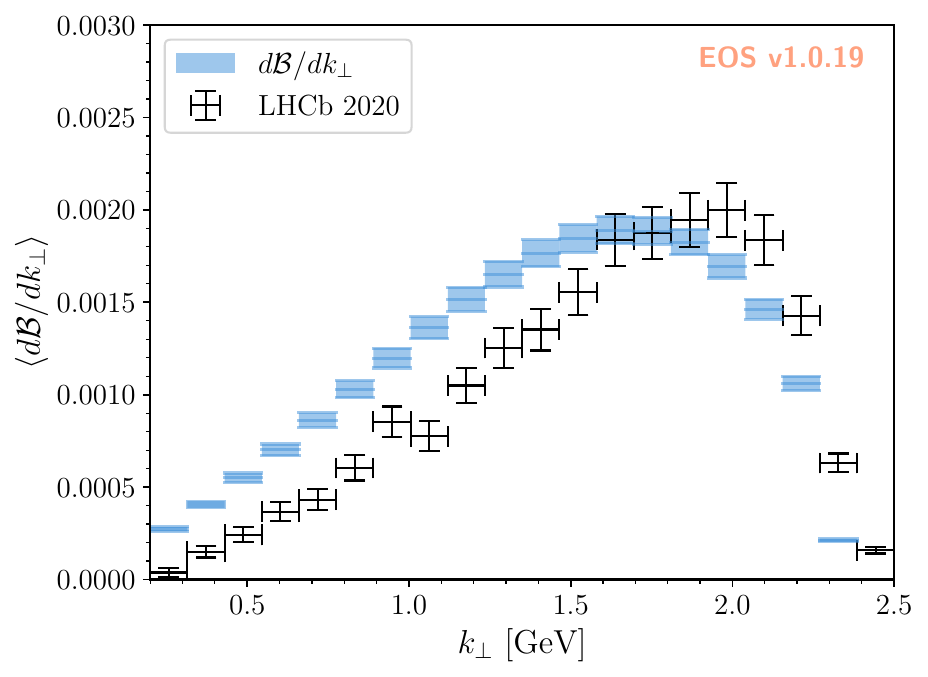}
    \caption{%
        The rate of $\bar{B}_s^0\to D_s^+\mu^-\bar\nu$ as measured
        by the LHCb experiment~\cite{LHCb:2020cyw} (black error bars
        labelled as LHCb 2020), which is still affected by some detector effects, overlaid 
        with our theoretical prediction \cref{eq:theoretical-predictions:differential-br-kperp2}
        in bins of \kperp (shown as blue bands), which does not account for any detector effects.
    }
    \label{fig:theory-vs-reconstructed}
\end{figure}
The result shown in \cref{eq:theoretical-predictions:differential-br-kperp2} illustrates that,
while we can predict the rate in $\kperp^2$, we cannot analytically invert the relationship
between the $\kperp^2$ rate and the $q^2$ rate, since it depends on the integration with respect
to an inaccessible kinematic variable. However, we can still predict the $\kperp^2$ rate through
numerical integration of the $q^2$ rate, which in turn enables us to infer information through
forward modelling. The predicted $\kperp$ rate is shown in \cref{fig:theory-vs-reconstructed},
overlaid with the detector-level results from the LHCb measurement.
Our prediction includes theoretical uncertainties obtained
from lattice QCD predictions for the form factors $f_+(q^2)$ and $f_0(q^2)$ by the HPQCD collaboration~\cite{McLean:2019qcx}
and accounts for the mass of the muon.
The data points of the measurement are obtained by extracting the yields and the uncertainties in each \kperp bin from
Fig.~6a from the supplementary material in Ref~\cite{LHCb-2019-041-CDS},
and applying Eqs.~(34)--(36) of Ref.~\cite{LHCb:2020cyw} to obtain the branching ratios.
For this measurement, LHCb normalises to the branching ratio of the reference decay $\bar{B}\to D\mu^-\bar\nu$,
leading to partial cancellation of the efficiencies in the ratio of the signal and reference decay.
As \cref{fig:theory-vs-reconstructed} shows, the agreement between the experimental results and our prediction is very poor,
indicating substantial detector effects that need to be taken into account.

\section{Forward modelling of detector effects}
\label{sec:detector-effects}

\subsection{Resolution and efficiency}

The geometry of the detector, the electromagnetic fields within the detector, the material of the detector components, and the loss of energy of charged particles due to bremsstrahlung, the event selection and the detector-specific trigger, all lead to a mismatch between the predictions made
at the theory level (labelled as ``th'') and the measurements made at the detector level (labelled as ``det'').
To account for these effects, one needs to model the relation between the theory level and the detector level.
Modelling in the direction from theory to detector level can be accomplished by detector-specific Monte Carlo simulations; it is commonly understood as \textit{forward modelling}. Achieving the reverse is considerably more difficult~\cite[Chapter 6]{Behnke:2013pga}.
The forward modelling is commonly expressed as the convolution
\begin{equation}
    \label{eq:detector-effects:convolution}
    P^\text{det}(\kperpdet)
        = \iint d\kperpdet\, d\kperpth\, A(\kperpth, \kperpth - \kperpdet)\, P^\text{th}(\kperpth)\,,
\end{equation}
In the above, $P^\text{th}$ is a theory-level probability density (PDF)
describing the distribution of events as a function of \kperpth,
$P^\text{det}$ is a density (but not necessarily a PDF) describing the detected events,
and the acceptance
\begin{equation}
    A(\kperpth, \kperpth - \kperpdet)
        = \varepsilon(\kperpth) \, r(\kperpth - \kperpdet)
\end{equation}
is the product of the detector efficiency function $\varepsilon$
and the detector resolution distribution $r$.
One can generalise \cref{eq:detector-effects:convolution} by replacing
\begin{equation}
    \begin{aligned}
        P^\text{th}(\kperpth)
            & \to \frac{d\mathcal{B}^\text{th}}{d\kperpth}\,,
            &
        P^\text{det}(\kperpdet)
            & \to \frac{d\mathcal{B}^\text{det}}{d\kperpdet}\,.
    \end{aligned}
\end{equation}
In the following, we will approximate the detector efficiency
and the detector resolution from Fig.~7a and Fig.~7c from the supplementary material~\cite{LHCb-2019-041-CDS}
provided as part of Ref.~\cite{LHCb:2020cyw}.
As a peculiarity of the setup of the analysis in Ref.~\cite{LHCb:2020cyw},
the average $\langle\epsilon\rangle$ of the efficiency $\epsilon$ has already been accounted for in the
published results to cancel systematic uncertainties associated with the normalisation mode,
\begin{equation}
    \langle\epsilon\rangle
        \equiv
            \frac{1}{\kperp|_\text{max} - \kperp|_\text{min}}
            \int_{\kperp|_\text{min}}^{\kperp|_\text{max}} d\kperp \epsilon(\kperp)\,.
\end{equation}
Since this step cannot be undone,
we make the replacement $\epsilon \to \hat\epsilon = \epsilon / \langle \epsilon\rangle$ whenever we apply the forward-modelling procedure in this analysis.
We advise caution that the results shown below are approximate and should be used with care, since
they do not (and cannot) account for either the statistical uncertainties due to the LHCb
Monte Carlo simulations or the systematic uncertainties (e.g., due to our choices of fit models).
A complete forward modelling framework, accounting for all systematic uncertainties, can only be provided by the LHCb collaboration.

\begin{figure}
    \centering
    \includegraphics[width=0.495\linewidth]{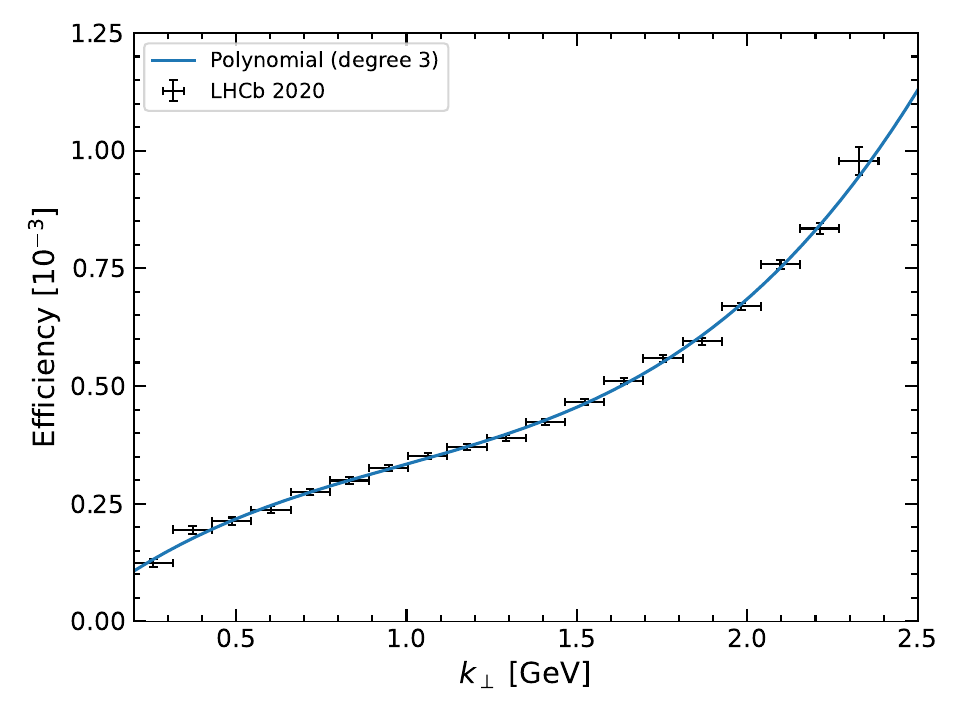}
    \includegraphics[width=0.495\linewidth]{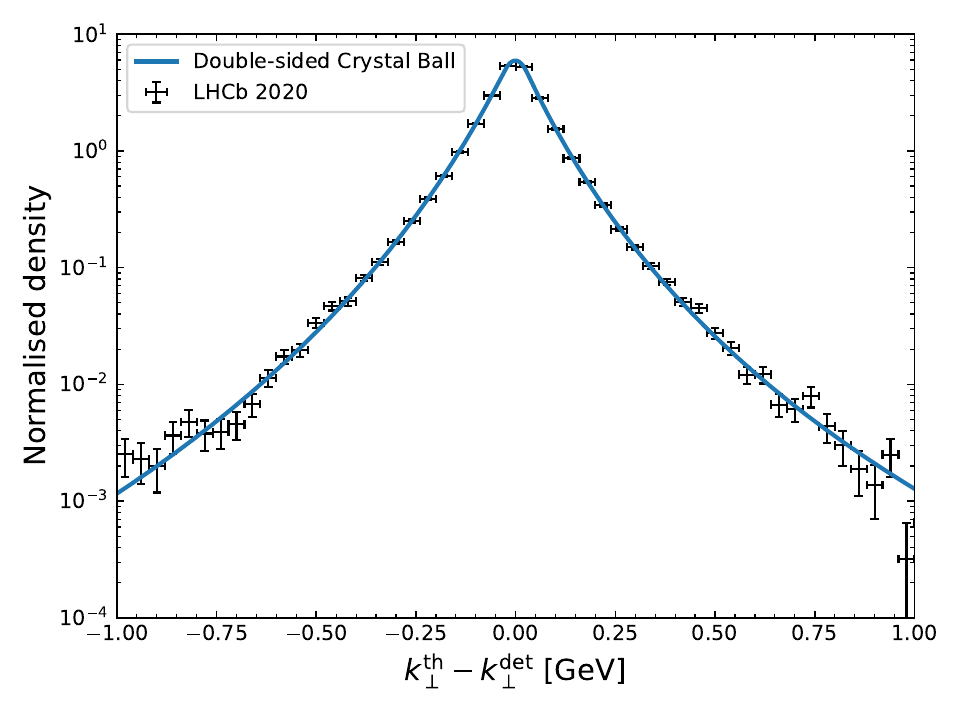}
    \caption{%
        The efficiency (left-hand side) and resolution (right-hand side) as documented in the supplementary material~\cite{LHCb-2019-041-CDS}
        of the LHCb measurement~\cite{LHCb:2020cyw} (black error bars). We fit the efficiency with a Legendre polynomial of degree three and the resolution with a double-sided Crystal Ball density (blue curves).
    }
    \label{fig:detector-effects:efficiency-resolution}
\end{figure}

\paragraph{Efficiency $\boldsymbol{\epsilon}$}
The efficiency determined by LHCb by means of Monte Carlo simulation is shown in Fig.~7a of the supplementary material~\cite{LHCb-2019-041-CDS} in the form of 19 bins\footnote{%
    The efficiency is presented in the same binning scheme as the data. However,
    bin \#20 is not included in the histogram that is part of the supplementary
    material~\cite{LHCb-2019-041-CDS}.
} in \kperp.
We model the efficiency function $\epsilon$ on the interval $\kperp \in [0.2, 2.5]\,\text{GeV}$ in terms of Legendre polynomials $P_n$ up to degree three,
\begin{equation}
    \label{eq:detector-effects:efficiency-model}
    \epsilon(\kperp) = \frac{d\zeta}{d\kperp} \sum_{n=0}^{n=3} \epsilon_n P_n(\zeta)\,,
    \qquad
    \zeta \equiv 2\frac{\kperp - \kperp|_\text{min}}{\kperp|_\text{max} - \kperp|_\text{min}} - 1\,.
\end{equation}
A log-likelihood is constructed by matching the continuous efficiency model \cref{eq:detector-effects:efficiency-model}
to the reported efficiency values and uncertainties, assuming them to be located at the bin centres.
The model parameters are fitted by maximising this likelihood with the best-fit point obtained as
\begin{equation}
    \begin{aligned}
        \epsilon_0
            & = 5.53 \cdot 10^{-4}\,,
            &
        \epsilon_1
            & = 4.90 \cdot 10^{-4}\,,
            &
        \epsilon_2
            & = 1.58 \cdot 10^{-4}\,,
            &
        \epsilon_3
            & = 9.74 \cdot 10^{-5}\,.
    \end{aligned}
\end{equation}
At the best-fit point, we find $\chi^2 = 19.35$, which corresponds to $p = 19.8\%$ assuming
$15$ degrees of freedom. This indicates a good fit quality. The result of our approximation of the efficiency
is shown in the left-hand side of \cref{fig:detector-effects:efficiency-resolution}.
The normalised efficiency $\hat\epsilon$ is obtained from the above as
\begin{equation}
    \hat\epsilon(\kperp) = \frac{1}{2}\left[1 + \sum_{n=1}^{n=3} \frac{\epsilon_n}{\epsilon_0} P_n(\zeta)\right]\,.
\end{equation}
\\
\paragraph{Resolution $\boldsymbol{r}$}
The resolution determined by LHCb by means of Monte Carlo simulation is shown in Fig.~7c of the
supplementary material~\cite{LHCb-2019-041-CDS} in form of 60 bins in $\Delta_{k_\perp} = \kperpth - \kperpdet$.
We describe the resolution through a Double-Sided Crystal Ball (DSCB) function,
\begin{equation}
    \label{eq:detector-effects:resolution-model}
    \text{DSCB}(\Delta_{k_\perp}; \mu, \sigma, \alpha_i, n_i) = \begin{cases} 
      a_L\left( b_L - \frac{\Delta_{k_\perp} - \mu}{\sigma}\right)^{-n_L} & \text{for } \frac{\Delta_{k_\perp}-\mu}{\sigma}\leq -\alpha_L, \\
      \exp\left\{-\frac{\left(\Delta_{k_\perp}-\mu\right)^2}{2\sigma^2}\right\} & \text{for }-\alpha_L < \frac{\Delta_{k_\perp}-\mu}{\sigma} \leq \alpha_R, \\
      a_R\left( b_R + \frac{\Delta_{k_\perp} - \mu}{\sigma}\right)^{-n_R} & \text{for } \alpha_R<\frac{\Delta_{k_\perp}-\mu}{\sigma},
   \end{cases}
\end{equation}
where $\alpha_i>0$ $(i = L,R)$ and
\begin{align}
        a_i
            & = \left(\frac{n_i}{\alpha_i}\right)^{n_i}\exp\left\{-\frac{\alpha_i^2}{2}\right\}\,,
            &
        b_i
            & = \frac{n_i}{\alpha_i} - \alpha_i.
\end{align}
We construct a log-likelihood by matching the continuous resolution model \cref{eq:detector-effects:resolution-model}
to the reported resolution values, assuming them to be located at the bin centres.
The uncertainties on the bin contents are not provided in the LHCb measurement. We assume a Poisson distribution
for the content $r_i$ of the $i$th bin and therefore approximate its uncertainties as $\sqrt{r_i}$.
This procedure forces us to remove one bin from the fit, since its bin content is zero.
Maximising this likelihood comprised of the remaining $59$ bins, we obtain the best-fit point as
\begin{equation}
    \begin{aligned}
        \mu
            & = -0.08 \cdot\ 10^{-2} \,,
            &
        \alpha_L
            & = 0.665\,,
            &
        n_L
            & = 7.590\,,
            \\
        \sigma
            & = 4.27 \cdot 10^{-2}\,,
            &
        \alpha_R
            & = 0.718\,,
            &
        n_R
            & = 6.558\,.
    \end{aligned}
\end{equation}
At the best-fit point, we find $\chi^2 = 67.247 $, which corresponds to $p = 9\%$
assuming $53$ degrees of freedom. This indicates a good-quality fit.

\subsection{Approximate response matrix}
\label{sec:detector-effects:response-matrix}

\begin{figure}
    \centering
    \includegraphics[width=0.495\linewidth]{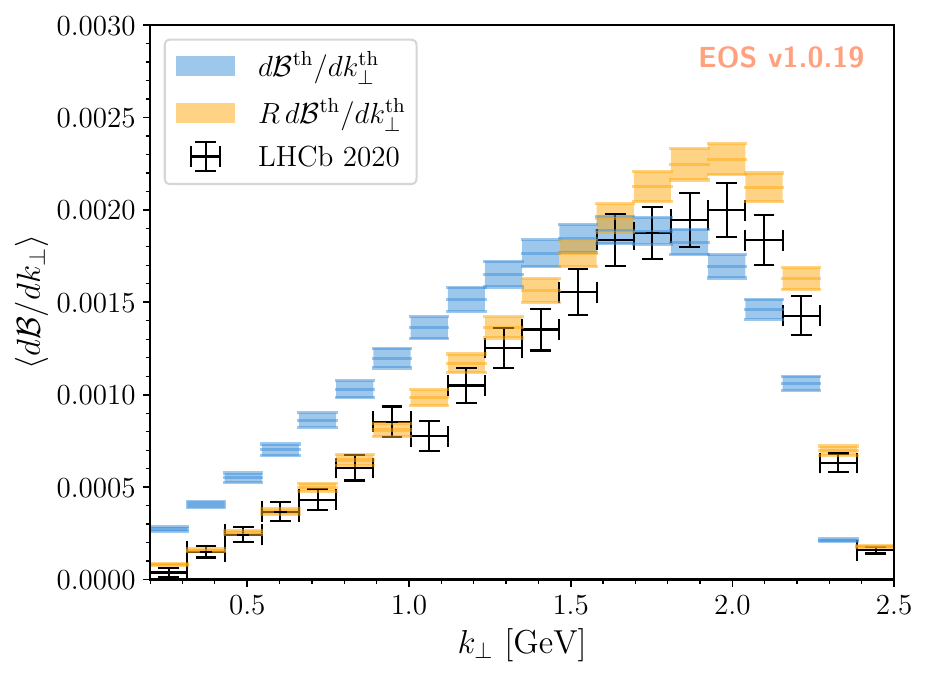}
    \includegraphics[width=0.495\linewidth]{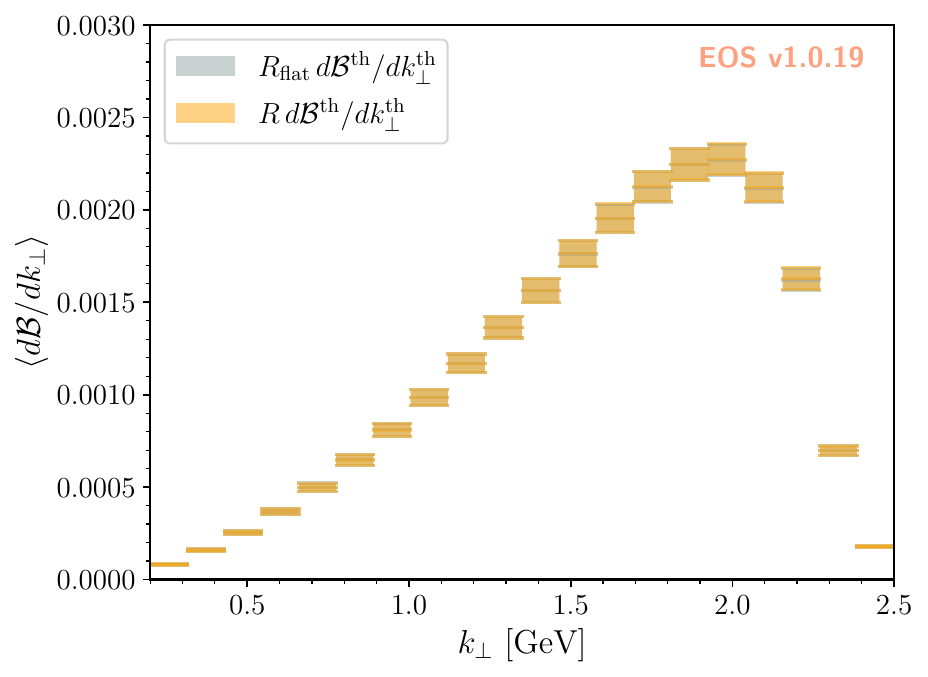}
    \caption{%
        The left-hand plot shows the rate of $\bar{B}_s^0\to D_s^+\mu^-\bar\nu$ as obtained
        from the LHCb experiment ~\cite{LHCb:2020cyw} including detector effects (black error bars,
        labelled LHCb 2020).
        It is overlaid 
        with our theory-level prediction \cref{eq:theoretical-predictions:differential-br-kperp2}
        in bins of \kperp (blue bands)
        and our detector level prediction obtained by multiplication
        with the response matrix \cref{eq:detector-effects:response-matrix:def}
        obtained from the full acceptance function $A$ (orange bands).
        The right-hand plot shows the lack of variability of the forward folding
        by plotting the detector-level effect using two different
        response matrices $R^\text{flat}$ (grey bands) and $R$ (orange bands),
        which are visually indistinguishable.
        The lack of variation of the response matrix indicates little to no dependence
        on the underlying signal model.
    }
    \label{fig:detector-effects:comparison}
\end{figure}

Since the data published in Ref.~\cite{LHCb:2020cyw} is provided
in bins of $\kperpdet$, we cannot apply \cref{eq:detector-effects:convolution} as is.
Instead, we apply the same binning scheme as used in Ref.~\cite{LHCb:2020cyw}, which provides for 20 equally-sized bins
in the interval $0.2\,\GeV \leq \kperpdet \leq 2.5\,\GeV$,
\begin{equation}
    P_m^\text{det}
        \equiv \int_{\text{bin}\,m} d\kperpdet P_\text{det}(\kperpdet)\,.
\end{equation}
In our aim to approximate the relation between theory and detector-level
quantities, we apply the same binning scheme to the theory predictions
expressed in terms of \kperpth,
\begin{equation}
    P_n^\text{th}
        \equiv \int_{\text{bin}\,n} d\kperpth P_\text{th}(\kperpth)\,.
\end{equation}
In the following the \textit{response matrix} $R_{mn}$ is determined, which describes the system of linear equations
\begin{equation}
    \label{eq:detector-effects:response-matrix:def}
    P_m^\text{det} = \sum_{n} R_{mn} P_n^\text{th}\,.
\end{equation}
For this, we compute
\begin{equation}
    \label{eq:bin-n-response-component}
    P_m^\text{det}\big|_{n}
        \equiv \int_{\text{bin}\,m} d\kperpdet \int_{\text{bin}\,n} d\kperpth A(\kperpth, \kperpth - \kperpdet) P^\text{th}(\kperpth)\,,
\end{equation}
and identify
\begin{equation}
    \label{eq:eesponse-matrix-def}
    R_{mn} = \frac{P_m^\text{det}\big|_{n}}{P_n^\text{th}}\,.
\end{equation}
We provide a machine-readable representation of the response matrix as part of our supplementary material~\cite{EOS-DATA-2025-06}.
The effect of applying the response matrix $R_{mn}$ is illustrated in \cref{fig:detector-effects:comparison}.
After this application, we find good agreement between our detector-level predictions and the data from Ref.~\cite{LHCb:2020cyw}.
We abstain from a quantitative comparison at this point, preferring to carry out a
full phenomenological analysis that includes variation of the hadronic form factors and $|V_{cb}|$ in \cref{sec:pheno}.
However, we briefly investigate how strongly the response matrix depends on the signal shape $P^\text{th}$,
assuming that the detector acceptance (\ie, efficiency $\epsilon$ and resolution $r$) has no or negligible dependence on
the signal shape.
To this end, a new response matrix $R^\text{flat}_{mn}$ is defined.
It is obtained by replacing the decay-specific signal shape $P^\text{th}$
with a uniform distribution in the kinematic variable $q^2$ defined over the interval $q^2_\mathrm{min}=0 \leq q^2 \leq q^2_\mathrm{max}=(M_{B_s}-M_{D_s})^2$,
\begin{equation}
    P^\mathrm{flat}_\mathrm{th}(q^2)  = \frac{1}{q^2_\mathrm{max}-q^2_\mathrm{min}}\Theta(q^2-q^2_\mathrm{min})\Theta(q^2_\mathrm{max}-q^2)\,.
\end{equation}
After applying the same change of variable $(q^2,\, \cos\theta_B) \mapsto (k_\perp,\, \cos{\theta_B})$ as discussed in \cref{sec:theoretical-predictions}
to this new (flat) signal shape, we obtain
\begin{equation}
    P^\mathrm{flat}_\mathrm{th}(k_\perp) = \frac{k_\perp}{q^2_\mathrm{max}-q^2_\mathrm{min}}\int\frac{M_{B_s}\,d\cos{\theta_B}}{(1-\cos^2\theta_B)\sqrt{M^2_{D_s}+\frac{k^2_\perp}{1-\cos^2\theta_B}}}\,.
\end{equation}
We compute the model-agnostic response matrix $R^\mathrm{flat}_{mn}$ using \cref{eq:bin-n-response-component,eq:eesponse-matrix-def};
it is also provided as part of the supplementary material~\cite{EOS-DATA-2025-06}.
Applying this response matrix $R^\mathrm{flat}$ to our theory-level predictions yields results that are visually indistinguishable
from those obtained by applying $R$, the response matrix obtained by using the signal model. Beyond visual agreement, we find that the eigenvalues of both matrices all agree within less than $0.4\%$.
This finding is one of the main results of this work.
Assuming (as stated earlier) that neither $\epsilon$ nor $r$ depend implicitly on the signal shape,
it indicates that interpretations of the measurements of the perpendicular momentum distribution
of semileptonic decays are stable against mismodelling.
Such mismodelling is possible due to incomplete knowledge of the hadronic form factors, or due to potential BSM effects in semileptonic decays,
or both.
As such, experimental measurements of the perpendicular momentum distribution would serve as important cross-checks of the extant
discrepancies found~\cite{Martinelli:2023fwm,Bordone:2024weh,Bordone:2025jur}
between the shape of the hadronic form factors in semileptonic $\bar{B}\to D^*$ transitions across
lattice QCD determinations~\cite{FermilabLattice:2021cdg,Harrison:2023dzh,Aoki:2023qpa}
and experimental measurements~\cite{Belle-II:2018jsg,Belle:2023bwv,Belle-II:2023okj}.
Put differently, the interpretation of measurements of the perpendicular momentum distribution is not contingent on the use of sophisticated software such
as \texttt{HAMMER}~\cite{Bernlochner:2020tfi} or \texttt{redist}~\cite{redist:v1.0.4} to ensure correctness of the result when changing the signal model.

\section{Phenomenological analysis}
\label{sec:pheno}

With the data and the response matrix at hand, we perform a Bayesian analysis using the available LHCb data~\cite{LHCb:2020cyw}.
For this analysis, we assume SM-like dynamics to infer the value of $|V_{cb}|$ from the data and to
challenge theoretical predictions for the $\bar{B}_s^0\to D_s^+$ form factors~\cite{McLean:2019qcx}.
We implement the theoretical predictions in bins of \kperp in the
\EOS software~\cite{EOSAuthors:2021xpv} as of version 1.0.19~\cite{EOS:v1.0.19}.
To draw samples from the posterior distributions, \EOS uses \dynesty~\cite{dynesty:v2.0.3},
an open-source software that implements dynamical nested sampling~\cite{Higson:2018}.
We provide our \EOS analysis file and further files such as posterior samples
and figures as part of the supplementary material~\cite{EOS-DATA-2025-06}.

\paragraph{Statistical model and choice of priors}\label{sec:pheno:priors}
Since we assume SM-like dynamics in the decay $\bar{B}_s^0\to D_s^+\ell^-\bar\nu$,
we impose a purely left-handed EFT coupling. We further account
for hard electro-magnetic corrections by means of the Sirlin factor~\cite{Sirlin:1980nh}.
In this setup, we encounter one free parameter in the Lagrangian, which corresponds to $|V_{cb}|$.
In fits that determine $|V_{cb}|$, we use a uniform prior PDF over the interval $[30, 50]\cdot 10^{-3}$.
\\

When dealing with the hadronic matrix elements for $\bar{B}_s^0\to D_s^+$ transitions, we
have to parametrise two hadronic form factors: the vector form factor $f_+(q^2)$, and the
scalar form factor $f_0(q^2)$. Each is parametrised in a truncated, simplified series
expansion in the complex $z$ plane, following Ref.~\cite{Bharucha:2015bzk},
\begin{equation}
    \begin{aligned}
        f_+(q^2)
            & = \frac{1}{1 - q^2 / M_{B_s^*}^2} \sum_{k=0}^{K} \alpha^{(+)}_k \left[z(q^2) - z(0)\right]^k\,,\\
        f_0(q^2)
            & = \frac{1}{1 - q^2 / M_{B_{s,0}}^2} \sum_{k=0}^{K} \alpha^{(0)}_k \left[z(q^2) - z(0)\right]^k\,.
    \end{aligned}
\end{equation}
In the above, we use the conformal map from the complex $q^2$ plane to the unit disk in $z$,
\begin{equation}
    z(q^2) \equiv \frac{\sqrt{t_+ - q^2} - \sqrt{t_+ - t_0^{\phantom{2}}}}{\sqrt{t_+ - q^2} + \sqrt{t_+ - t_0^{\phantom{2}}}}\,,
\end{equation}
where we abbreviate $t_\pm \equiv (M_{B_s} \pm M_{D_s})^2$ and $t_0 \equiv t_+ (1 - \sqrt{1 - t_-/t_+})$.
The masses of the first one-body intermediate states emerging in each form factor are fixed to be $M_{B_s^*} = 5.415\,\GeV$
and $M_{B_{s,0}} = 5.630\,\GeV$, respectively.
The identity $f_+(0) = f_0(0)$ is used to replace the expansion coefficient $a^{(0)}_0$ with a linear
combination of the remaining expansion coefficients.
With truncation of the series in $z$ at order $K = 2$, the form factors
contribute a total of five hadronic parameters to our analysis.
We use uniform priors for these parameters, chosen wide enough so as not to cut into
the peak resulting from the likelihood.
The concrete ranges are documented as part of the
supplementary material~\cite{EOS-DATA-2025-06} as priors \texttt{FF} \& \texttt{FF-Projection}.
The latter is needed to increase the range for the parameter $\alpha_2^{(+)}$ when fitting to projected LHCb results,
which are defined below.

\paragraph{Likelihood}\label{sec:pheno:likelihood}
We use lattice QCD results by the HPQCD collaboration for the two form factors $f_+$ and $f_0$~\cite{McLean:2019qcx}.
These results are incorporated in form of a theoretical five-dimensional multivariate Gaussian likelihood of the $f_+$ form factor in three $q^2$ points
and the $f_0$ form factor in two $q^2$ points. This likelihood (labelled ``LQCD'' below) is available in \EOS via the constraint
\begin{center}
    \texttt{B\_s->D\_s::f\_++f\_0@HPQCD:2019A}
\end{center}

The LHCb analysis~\cite{LHCb:2020cyw} provides the detector-level yields of the kinematic distribution of the decay $\bar{B}_s^0\to D_s^+\mu^-\bar\nu$
in 20 bins of \kperp. As discussed at the end of \cref{sec:theoretical-predictions}, Ref.~\cite{LHCb:2020cyw} provides sufficient information
to convert these yields into binned branching ratio measurements. Through the normalisation of the yields in terms of a reference channel,
the average efficiency $\langle \epsilon\rangle$ is already accounted for.
Due to lack of correlation information across the bins, we are forced to treat the individual measurements as 20 uncorrelated
Gaussian likelihoods. However, for convenience, we use the measurement within a single 20-dimensional multivariate Gaussian
likelihood with a diagonal covariance matrix. This setup permits us to transparently apply the forward modelling discussed
in \cref{sec:detector-effects}. All information entering this likelihood (labelled ``data'' below), including the response matrix, are contained in
the \EOS constraint named
\begin{center}
    \texttt{B\_s\^{}0->D\_s\^{}+mu\^{}-nu::KinematicalDistribution[kperp,response]@{}LHCb:2020C}
\end{center}
which is recorded in the analysis file within the supplementary material~\cite{EOS-DATA-2025-06}.
To investigate future performance of this type of measurement, we repeat our analysis with a modified experimental likelihood.
The constraint used to construct this second likelihood (labelled ``projection'' below),
\begin{center}
    \texttt{B\_s\^{}0->D\_s\^{}+mu\^{}-nu::KinematicalDistribution[kperp,response]@{}Projection}
\end{center}
differs from the first only by a rescaling of the covariance matrix by a factor $1/10$.
This rescaling corresponds to the expectation by LHCb to obtain roughly ten times as many samples of
the decay $\bar{B}_s\to D_s\mu^-\bar\nu$ as used in Ref.~\cite{LHCb:2020cyw} after the completion of
LHCb Run 3 data-taking (2022-2026)~\cite{RevModPhys.94.015003}.

\begin{figure}[t]
    \centering
    \includegraphics[width=.66\linewidth]{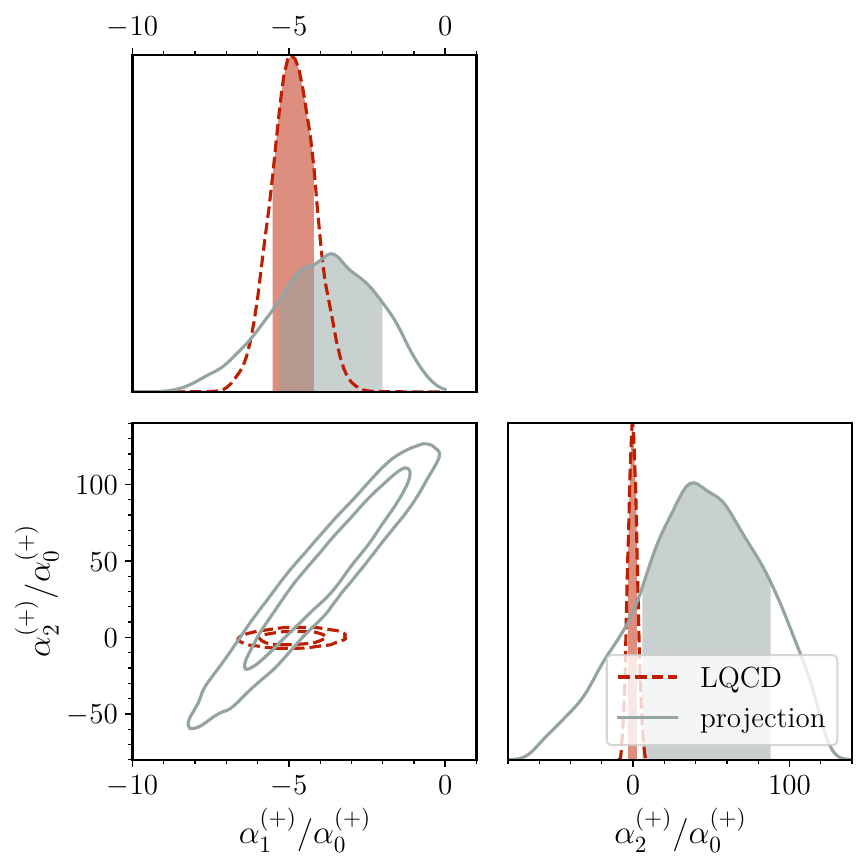}
    \caption{%
        The one-dimensional marginal posterior distributions the ratios of form factor
        parameters $\alpha_1^{(+)}/\alpha_0^{(+)}$ and $\alpha_2^{(+)}/\alpha_0^{(+)}$,
        accompanied by their joint two-dimensional marginal posterior distribution.
        Curves show the posterior density, and shared areas/contours indicate the central interval/region
        at $68\%$ probability
        (theory likelihood: grey curves, areas, and contours; LHCb Run-3 projection: dashed red curves, areas, and contours).
    }
    \label{fig:pheno:results}
\end{figure}
\paragraph{Results}\label{sec:pheno:results}
We find good agreement between the data and our posterior predictions for the \kperp-binned
branching ratio. This is quantified by determining the $\chi^2$ value at the best-fit point. We obtain $\chi^2 = 8.77$
for $19$ degrees of freedom. This corresponds to a $p$ value of $\sim 97\,\%$,
indicating excellent agreement between the theory prediction and the data.
We obtain
\begin{equation}
    \label{eq:pheno:results:Vcb}
    |V_{cb}| = 38.60^{+0.81}_{-0.80} \times 10^{-3}\,.
\end{equation}
Our result agrees with the one obtained by the LHCb collaboration at slightly within 2 standard
deviations. We stress, however, that this comparison is not very meaningful, given that this determination
of $|V_{cb}|$ depends overwhelmingly on the normalisation of the LHCb measurement.
The precision of this normalisation, the absolute branching fraction of $\bar{B}\to D\mu^-\bar\nu$ decays,
sets a lower bound on a systematic error of the extraction reported in \cref{eq:pheno:results:Vcb}.
Beside $|V_{cb}|$, we determine the ratios of form factor parameters
$\alpha_1^{(+)} / \alpha_0^{(+)}$ and $\alpha_2^{(+)} / \alpha_0^{(+)}$.
We find that the posterior distribution for $\alpha_1^{(+)} / \alpha_0^{(+)}$ is slightly narrower
and shifted with respect to its prior distribution, indicating that additional information was
inferred from the distribution of the branching ratio in $\kperp$. We do not find a similar
effect in either $\alpha_0^{(+)}$ or $\alpha_2^{(+)} / \alpha_0^{(+)}$.
Nevertheless, knowledge of the form factors can successfully be inferred
from the \kperp distribution without relying on internal LHCb knowledge. This enables us and other
groups to utilise this type of measurement in global analyses.

In addition, we perform an analysis that includes the ``projection'' likelihood but does not
include the ``LQCD'' likelihood. With this analysis, we investigate how the information
contained in the projected results for the \kperp rate compares to the already available information
from lattice QCD. Fixing $|V_{cb}| = 39\times 10^{-3}$, we obtain the marginal posterior distributions
for the ratios $\alpha_1^{(+)} / \alpha_0^{(+)}$ and $\alpha_2^{(+)} / \alpha_0^{(+)}$ shown
in \cref{fig:pheno:results}. We note that for this scenario the results for $\alpha_1^{(+)} / \alpha_0^{(+)}$
from the projected LHCb data only start to become competitive with the lattice QCD results.
Moreover, as can be seen in the 2D marginal posterior, the lattice QCD results and the projected
dataset provide complementary constraints on the parameter space.
Nevertheless, the sensitivity to the ratio $\alpha_2^{(+)} / \alpha_0^{(+)}$ remains poor.
We suggest to investigate a non-uniform
binning scheme for the \kperp measurement to maximise the amount of information that can be inferred.

\section{Summary}
\label{sec:summary}

We have investigated how to use the kinematic distribution of semileptonic decays with respect
to the perpendicular projection \kperp of the momentum of the final state hadron $k$.
The theoretical predictions for the differential \kperp rate are obtained in a straight-forward way from the decay distribution
in the more common variable $q^2$, the square of the four-momentum of the lepton-neutrino system.
Our results suggest that the \kperp distribution can only be obtained from forward-modelling,
due to the dependence of a secondary hidden kinematic variable that we label $\cos\theta_B$.
Crucial input is the probability density for $\cos\theta_B$, which factorises from the $q^2$ density
in the rest frame for a (pseudo)scalar initial state.
The generalisation to other initial states seems possible but more intricate.
\\

We have applied our theoretical results to the available experimental data on the \kperp distribution
of $\bar{B}_s^0\to D_s^+\mu^-\bar\nu$ decays, as a case study. This data is still affected
by a subset of the identified detector effects.
Effective use of this data thus involves forward modelling of the remaining detector effects.
To this end, we fit the efficiency and detector resolution functions.
We use the fits of these functions to compute a response matrix $R$, which is used to transform
our theory-level predictions to the detector level. The main result of this part of our analysis
is the observation that the response matrix is stable against strong variations of the signal model
and therefore stable against mismodelling of the signal due to, \eg, mismodelling of the underlying
hadronic form factors or due to potential BSM effects. We provide the response matrices used
within our analysis as machine-readable files as part of our supplementary material.
We encourage the LHCb collaboration to corroborate our findings with their internal and more detailed
modelling of the detector effects.
\\

Finally, we have used our results to conduct a phenomenological analysis of the $\bar{B}_s^0\to D_s^+\mu^-\bar\nu$ decay
rate. We find excellent agreement between the data and our forward-modelled theoretical predictions,
with a $p$ value of $\sim 97\%$. We extract $|V_{cb}|$ and find our result to be in agreement with
the LHCb result at the $2\,\sigma$ level. However, we caution that this comparison is not meaningful
for two reasons. 
First, we do not use the same framework, \ie, the LHCb result is based on data beyond what is publicly accessible.
Second, the LHCb results are obtained from a simultaneous fit to measurements of the \kperp rate in $\bar{B}_s^0\to D_s^+\mu^-\bar\nu$
and the $k_{1,\perp}$ rate in $\bar{B}_s^0\to D_s^{*+}(\to D_s^+(k_1) \gamma(k_2))\mu^-\bar\nu$; we cannot use the latter in our analysis
as we discuss in \cref{app:other-variables}.
As the main result of our phenomenological analysis, we find that we can infer knowledge about the
$q^2$ shape purely from data on the binned \kperp rate.
\\

In conclusion, we recommend performing further measurements of semileptonic decay rates as functions of \kperp,
which will serve as important sources of shape information and as cross-checks of measurements that use the more
typical rates in $q^2$. However, such measurements should be accompanied by the necessary
information to account for detector effects in the forward modelling. We look forward to engaging with the experimental collaborations on this matter.

\acknowledgments

We thank Marcello Rotondo for a careful reading of the manuscript, and Mirco Dorigo, Martin Jung, and M\'eril Reboud for useful discussions.
DvD is grateful to the Mainz Institute for Theoretical Physics (MITP) of the Cluster of Excellence PRISMA$^+$
(Project ID 390831469), for its hospitality and its partial support during the first stages of this work.
CE and DvD acknowledge support by the UK Science and Technology Facilities Council (grant numbers ST/V003941/1 and ST/X003167/1).
BM acknowledges support by UK Science and Technology Facilities Council and the University of Manchester.

\appendix

\section{Generalisation to further variables fails}
\label{app:other-variables}

In the main matter, we show that in the $\bar{B}_s^0$ rest frame, the perpendicular momentum $\kperp$ of the $D_s^+$
is unambiguously related to the squared momentum $q^2$ by means of
\begin{equation}
    k_\perp^2 = \frac{\lambda(M^2_{B_s},\,M^2_{D_s},\,q^2)}{4M^2_{B_s}}\sin^2\theta_B\,.
\end{equation}
Put differently, the shape of the $\kperp$ distribution of the decay is predictably comprised of the shape
of the $q^2$ distribution and the shape of the $\cos\theta_B$ distribution.

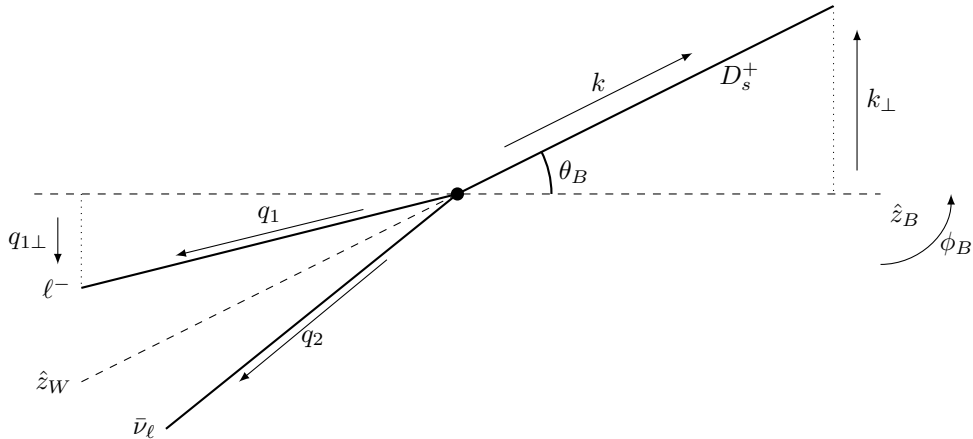
\begin{figure}[t]
    \centering
    \begin{tikzpicture}[scale=1.25]
        \fill (0,0) circle (0.7mm); 

        \draw[dashed] (-4.5,0) --  (4.5,0) node[below right] {$\hat{z}_B$};

        \draw[thick] (0,0) --  (4, 2) node[near end, below] {$D^+_{s}$};

        \draw[-latex] (0.5,0.5) --  (2.5,1.5) node[midway, above] {$k$};

        \draw[dotted] (4,2) -- (4,0);

        \draw[-latex] (4.25,0.25) -- (4.25,1.75) node[midway, right] {$k_\perp$};

        \draw[dashed] (0,0) -- (-4,-2) node[at end, left] {$\hat{z}_W$};
        
        \draw[dotted] (-4,0) -- (-4,-1);

        \draw[-latex] (-4.25,-0.25) -- (-4.25,-0.75) node[midway, left] {$q_{1\perp}$};

        \draw[thick] (0,0) -- (-4,-1) node[at end, left] {$\ell^-$};

        \draw[-latex] (-1,-0.165) -- (-3,-0.655) node[midway, above] {$q_1$};

        \draw[thick] (0,0) -- (-3.1,-2.5) node[at end, left] {$\bar{\nu}_\ell$};

        \draw[-latex] (-0.75,-0.7) -- (-2.325,-2) node[midway, below] {$q_2$};

        \draw[thick] (0,0) ++ (0:1) arc (0:27:1) node[midway, right] {$\theta_B$};
        \draw[-latex] (4.5,0) ++ (270:0.75) arc (270:360:0.75) node[midway, right] {$\phi_B$};
        
    \end{tikzpicture}
    \caption{%
        Sketch of the full decay kinematics and illustration of the $q_{1\perp}$ variable in
        the rest frame of the $\bar{B}_{s}^0$ meson.
        For the definition of the various quantities, see the caption of \cref{fig:intro:sketch}.
    }

    \label{fig:Secondary_Decay}
\end{figure}

Wondering if we can generalise this insight to other momenta, we turn to the charged-lepton momentum $q_1$, with the kinematics of the 
production mechanism
\begin{equation}
    W^*(q)\to\ell^-(q_1)\,\bar{\nu}_\ell(q_2)
\end{equation}
illustrated in \cref{fig:Secondary_Decay}.
Resolving the kinematics of the charged lepton, we gain access to one additional kinematic variable $\cos\theta_\ell$, which is related
to the scalar product $q_1 \cdot k$.
We are now interested in seeing if we can extract information on the joint $(q^2, \, \cos\theta_\ell)$ dependence from measurements
of the joined $(\kperp, q_{1\perp})$ dependence, where $q_{1\perp}$ is the perpendicular projection of the charged-lepton momentum $q_{1}$
on the $\bar{B}_s^0$ flight direction.
We find
\begin{equation}
    \begin{aligned}
        q_{1\perp}^2 & = \frac{q^2}{4}(1-\cos^2{\theta_\ell})\cos^2\theta_B + \frac{1}{4}\left(|\vec{k}|-\cos\theta_\ell\sqrt{|\vec{k}|^2+q^2}\right)^2\sin^2\theta_B \\
        & + \frac{\sqrt{q^2}}{2}\left(|\vec{k}|-\cos\theta_\ell\sqrt{|\vec{k}|^2+q^2}\right)\sqrt{1-\cos^2\theta_\ell}\sin\theta_B\cos\theta_B. \\
    \end{aligned}
\end{equation}
Qualitatively, this is a different result than for $\kperp$.
Critically, the dependence on the additional kinematic variable $\theta_\ell$ is nonlinear.
Attempting to invert this relation for $\cos\theta_\ell$ yields four distinct analytic branches
\begin{equation}
    \cos\theta_\ell : f_{s_1,s_2}(q^2,\theta_B,q_{1\perp})\,, \qquad s_1,s_2 \in \{+,-\}\,.
\end{equation}
For each of these branches, we find parts of the $(q^2,\theta_B,q_{1\perp})$ domain in which the branch yields a correct solution.
However, none of the branches yield the correct solution in the full domain.
Moreover, we cannot provide a piecewise-defined function comprised of all four solutions that is correct in the full domain.
As a consequence, we cannot unambiguously determine $\cos\theta_\ell$ for a given
value of $q_{1\perp}$, even if we assume to know $q^2$ and $\cos\theta_B$\,. This determination is critical to our approach,
and therefore our approach cannot be generalised to use the $q_{1\perp}$ dependence in the same way as the $\kperp$ dependence.
For decays to vector mesons (e.g., $D_s^{*+}(k) \to D_s^+(k_1) \gamma(k_2)$), we come to the same conclusion for the other side of the decay chain:
we cannot use $k_{1\perp}$, the projection of $k_1$ onto the direction perpendicular to the $\bar{B}_s^0$ meson direction of flight,
to extract information on the angular distribution in the $D_s^+ \gamma$ helicity angle or the azimuthal angle between the $\ell^-\bar\nu$
and $D_s^+\gamma$ decay planes.

\bibliographystyle{JHEP.bst}
\bibliography{references.bib}

\end{document}